# Distinguishing Local and non-Local Demagnetization in Ferromagnetic FePt Nanoparticles

Patrick W. Granitzka[1,2], Alexander H. Reid[3*], Jerome Hurst[4], Emmanuelle Jal[5], Loïc Le Guyader[1,6], Tian-Min Liu[1], Leandro Salemi[4], Daniel J. Higley[1,3], Tyler Chase[1], Zhao Chen[7], Marco Berritta[4], William F. Schlotter[3], Hendrik Ohldag[8], Georgi L. Dakovski[3], Sebastian Carron[3], Matthias C. Hoffmann[3], Jian Wang[9], Virat Mehta[10,11], Olav Hellwig[10,12,13], Eric E. Fullerton[14], Yukiko K. Takahashi[9], Joachim Stöhr[1], Peter M. Oppeneer[4] & Hermann A. Dürr[1,4#]

[1]*Stanford Institute for Materials and Energy Sciences, SLAC National Accelerator Laboratory, 2575 Sand Hill Road, Menlo Park, CA 94025, USA*
[2]*van der Waals-Zeeman Institute, University of Amsterdam, 1018XE Amsterdam, The Netherlands*
[3]*Linac Coherent Light Source, SLAC National Accelerator Laboratory, 2575 Sand Hill Road, Menlo Park, CA 94025, USA*
[4]*Department of Physics and Astronomy, Uppsala University, P. O. Box 516, S-75120 Uppsala, Sweden*
[5]*Sorbonne Universités, CNRS, Laboratoire de Chimie Physique – Matière et Rayonnement, 75005 Paris, France*
[6]*Spectroscopy and Coherent Scattering Instrument, European XFEL GmbH, Holzkoppel 4, 22869 Schenefeld, Germany*
[7]*Department of Physics, Stanford University, Stanford, California 94305, USA*
[8]*Stanford Synchrotron Radiation Light Source, SLAC National Accelerator Laboratory, 2575 Sand Hill Road, Menlo Park, CA 94025, USA*
[9]*Magnet Materials Unit, National Institute for Material Science, Tsukuba 305-0047, Japan*
[10]*San Jose Research Center, HGST a Western Digital Company, 3403 Yerba Buena Road, San Jose CA 95135, USA*
[11]*Thomas J. Watson Research Center, 1101 Kitchawan Road, Yorktown Heights, NY 10598, USA*
[12]*Institute of Physics, Technische Universität Chemnitz, Reichenhainer Straße 70, D-09107, Chemnitz, Germany*
[13]*Institute of Ion Beam Physics and Materials Research, Helmholtz-Zentrum Dresden–Rossendorf, 01328, Dresden, Germany*
[14]*Center for Memory and Recording Research, University of California San Diego, 9500 Gilman Drive, La Jolla, CA 92093-0401, USA*

Time-resolved coherent X-ray diffraction is used to measure the spatially resolved magnetization structure within FePt nanoparticles during laser-induced ultrafast demagnetization. The momentum-dependent X-ray magnetic diffraction shows that demagnetization proceeds at different rates at different X-ray momentum transfer. We show that the observed momentum-dependent scattering has the signature of inhomogeneous demagnetization within the nanoparticles, with the demagnetization proceeding more rapidly at the boundary of the nanoparticle. A *shell* region of reduced magnetization forms and moves inwards at a supermagnonic velocity. Spin-transport calculations show that the shell formation is driven by a superdiffusive spin flux mainly leaving the nanoparticle into the surrounding carbon. Quantifying this non-local contribution to the demagnetization allows us to separate it from the local demagnetization.

Magnetization dynamics is an intriguing multi-scale problem in condensed matter physics. Near equilibrium, the Landau-Lifshitz-Gilbert (LLG) equation describes the dynamics of magnetization down to nanosecond timescales and sub-micrometer length scales. However, future magnetic data technologies are pushing the development of magnetic elements with nanoscale dimensions that can be manipulated on sub-nanosecond timescales. It is in this range that the LLG description of magnetization begins to break down and new magnetic phenomena emerge. In particular, the electronic nature of spin transport becomes apparent on these length and timescales [1–9] as does highly efficient transfer of spin angular momentum to the lattice [10–14] and the ultrafast generation of magnons [15–17].

Magnetic spin transport at ballistic, superdiffusive or diffusive electronic velocities have all been shown to occur in nanomagnetic elements on subpicosecond timescales. Experimental realizations of these effects have focused on metallic heterostructures using non-magnetic metals either as a spin collector [1,4,7], or as a spin valve [5,18]. In these studies, comparisons between metal–metal and metal–insulator heterostructures are often used to quantify the effects of such hot-electron spin currents [6,7,19]. Spin current effects have also been observed in heterogenous alloy systems between nanometer scale regions [3]. Some recent experimental data also points to the importance of spin transport in homogenous magnetic elements [20–22]. However, separating the relative contributions of spin transport from local demagnetization has been a major challenge [6,23,24]. In addition, the role of boundary effects in ultrafast magnetic processes and the conservation of angular momentum has been recently highlighted [25]. How spin-current effects manifest in nanomagnetic elements with dimensions below the spin diffusion length remains unclear due to the difficulty in accessing this spatial resolution experimentally.

In this Letter we study the ultrafast loss of magnetization within FePt nanoparticles with nanometer length scale resolution and femtosecond timescale resolution using ultrashort resonant X-ray pulses. This new regime of magnetic transport includes the formation and propagation of a near ballistic demagnetization front from the edge of the nanoparticle inwards. The speed at which this region of *shell* demagnetization forms points to its origin: hot-electron spin transport. We quantify the size of the current flowing from the nanoparticle and discuss its transient behavior in terms of superdiffusive spin transport [2]. The results distinguish the homogenous bulk demagnetization effects from the inhomogeneous transport induced demagnetization.

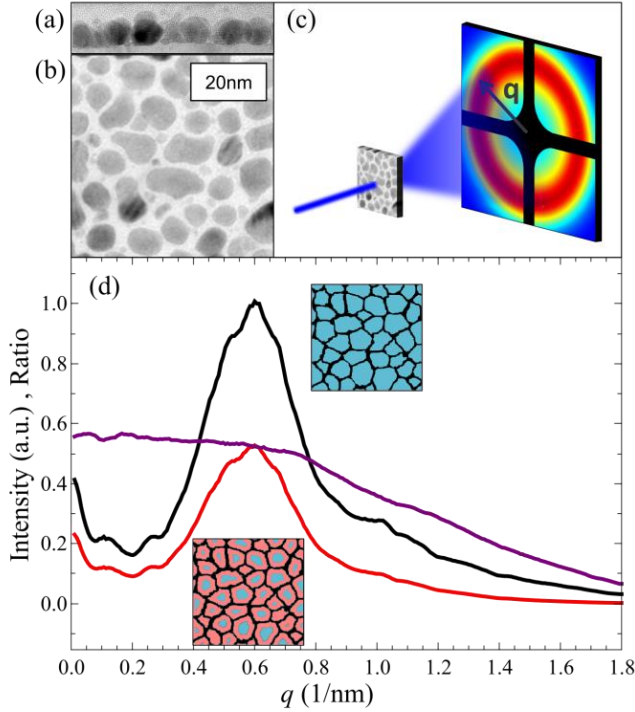

FIG. 1. Magnetic small angle scattering from an FePt array. (a) Cross-section view TEM image of FePt nanoparticles on MgO single crystal layer and surrounded by carbon. (b) Top (plan) view TEM image of FePt grains (gray) separated by carbon (white). (c) The experimental setup: an X-ray beam pulse (blue) passes through the FePt granular membrane sample. The diffraction pattern, caused by the different X-ray absorption of the FePt grains and the surrounding carbon, appears on the detector where it is digitized. (d) Shows simulated diffraction $I_{mag}(q)$ for homogeneously magnetized FePt nanoparticles (black), and for nanoparticles with a 50% demagnetized core and 75% demagnetized 2 nm thick shell (red). The ratio of the scattering intensities is shown in purple. The insets visualize the magnetic state of the different simulations.

The FePt–carbon film used in this study was grown epitaxially on a single-crystal MgO substrate which in turn was grown on a NiTa/TiN seed layer on a 100 nm SiN membrane film. The film is composed of single-crystal FePt nanoparticles in the $L1_0$ phase separated by an amorphous carbon segregant as shown in Fig. 1a.& 1b. The FePt nanoparticles form with their c-axis (the magnetic easy axis) normal to the film plane. The spatially resolved magnetism of the FePt nanoparticles was measured using circularly polarized resonant X-ray scattering on the peak of the Fe $L_3$ absorption edge in a transmission geometry (see Fig. 1c). Measurements of the magnetization dynamics were made using a pump–probe approach at the SXR hutch of the Linac Coherent Light Source. An 800 nm optical pulse of 50 fs duration and fluence 11 mJ/cm$^2$ was used as pump. X-ray pulses of 60 fs durations and 0.6 eV bandwidths were used as probe. Data was collected with continuously varied time delays at 120 Hz for right- & left-circularly polarized X-rays while applying an out-of-plane magnetic field of $\mu_0 H = \pm\, 0.4$ T to restore the magnetic state between measurements, see Fig. 1c. The X-ray–laser time of arrival was jitter corrected using an upstream monitor [26]. The X-ray diffraction signal was collected on an in-vacuum pnCCD camera positioned 100 mm behind the sample. The effects of inhomogeneous optical excitation of the FePt due to near-field interference were removed by making measurements for both magnetic field directions and extracting the magnetic component switching with the magnetic field (FePt nanoparticles where sufficient optical energy was absorbed to allow reversal) [27].

Resonant small-angle X-ray scattering allows the separation of the charge and magnetic characteristics in nano-magnetic materials [28–30]. At the Fe $L_3$ absorption resonance the scattering signal has two contributions in first order. An absorption scattering contribution ($C_q$) that is proportional to the contrast between the X-ray transmission through the FePt nanoparticles and the X-ray transmission through the surrounding carbon matrix, and a magnetic scattering contribution ($M_q$), that is proportional to the magnetic absorption contrast in the FePt nanoparticles due to the XMCD effect. At the Fe $L_3$ edge, the XMCD sum rules give that the magnetic scattering is proportional to a sum of the Fourier components of spin, S, and orbital, L, moments [12]:

$$M_q \propto S_q + 3/2 L_q.$$

The intensity recorded on the detector is proportional to the square of the absorption ($C_q$) and magnetic ($M_q$) scattering contributions:

$$I_{scat}(\sigma, B) = |C_q|^2 + |M_q|^2 \pm 2Re[C_q^* M_q],$$

where the sign $\pm$ depends on both the helicity of the light and the magnetic state of the sample [3,27]. We isolate the scattering intensity component of the switching nanoparticles, linear in $M_q$, by reversal of the external magnetic field upon excitation with the laser:

$$I_{mag} \equiv 4Re[C_q^* M_q] = (I_{\sigma_+, B_+} - I_{\sigma_-, B_+}) - (I_{\sigma_+, B_-} - I_{\sigma_-, B_-}),$$

where $\sigma_\pm$ is the X-ray helicity and $B_\pm$ is applied field direction [27].

To understand the momentum resolved magnetic scattering data we conducted simulations of the small-angle scattering from a model of the magnetic sample structure. The nanoparticle sample was modelled as an irregular array of FePt particles with fixed thickness; an SEM image is used as the basis of the structure. The X-ray optical and magnetic properties were defined over the model's FePt-C matrix with a grid size of 0.33 x 0.33 nm$^2$ (approximately the FePt unit cell size). Figure 1d shows the simulated magnetic scattering $I_{mag}(q)$ for the model structure. The fully magnetized magnetic scattering (Fig. 1d black curve) shows a broad peak as a function of the momentum transfer $q$ which is centered at the momentum associated with the average inter-grain separation. The magnetic properties of individual FePt nanoparticles can be defined in the simulation. In order to see how an intra-nanoparticle inhomogeneous magnetization can change the magnetic scattering, we have performed simulations for a *core-shell* magnetic structure, where a region at the boundary of the FePt nanoparticles (shell) has a lower magnetization ($0.25 M_s$) than the center (core) of the nanoparticle ($0.5 M_s$), see Fig. 1d red curve. This ratio of a demagnetized shell scattering is compared to the saturated magnetic scattering in Fig. 1d (purple line). It is observed that the shell demagnetized state shows a large reduction of the magnetic scattering primarily above the scattering peak when compared to the saturated state. This is the signature $q$-dependent change found in the experimental scattering data.

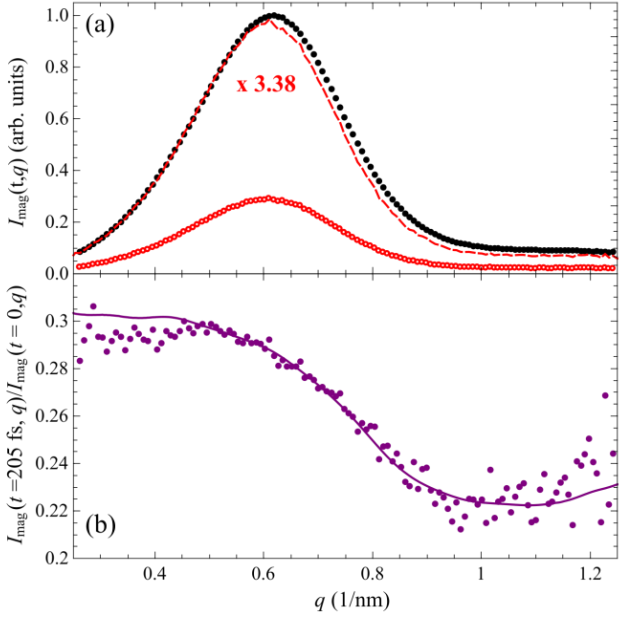

FIG. 2. Measured $I_{mag}(q)$ for demagnetized FePt nanoparticles. (a) Shows azimuthal integrated X-ray scattering dichroism $I_{mag}(q)$ for before the optical pump (black dots) and for 205 fs after optical pumping (red dots). The dashed red curve is the 205 fs data scaled by 3.38 times to match the before-pump data at $q = 0.5$ nm$^{-1}$. (b) Plots of $I_{mag}(q,t)/I_{mag}(q,0)$ for the experimental data (purple dots) and the best fit to these data for a simulated shell demagnetization (purple line).

The experimental magnetic scattering for before time zero (fully magnetized FePt) and for a time delay of 205 fs is shown in Fig. 2a (black and red dots resp.). This shows the same characteristics as the simulation with $I_{mag}(q)$ peaked at the $q$ corresponding to the inter-grain separation. These data further show that at 205 fs the magnetization of the FePt particles is reduced to approximately 30% of the initial value. To visualize the momentum dependent changes in the magnetic scattering we scale $I_{mag}(t = 205\text{ fs})$ so that it matches $I_{mag}(t = 0)$ at $q = 0.5$ nm$^{-1}$. The experimental data shows that there is a $q$-dependent change in the magnetic scattering profile, with the magnetic scattering at $q$ values above the peak reduced when compared those below the peak. This is the signature $q$-dependence of $I_{mag}(q)$ observed for shell demagnetized FePt simulation, albeit, to a lesser degree of shell demagnetization. We note that investigations of the magnetic scattering showed that a boundary reduction in magnetization was the only magnetic change that could lead to the observed $q$-dependent change in magnetic scattering and that effects such as particle-size dependent demagnetization had distinctly different $q$-dependent signatures.

To analyze the spatial changes in the magnetic scattering during the demagnetization process we consider the normalized $q$-dependent magnetic scattering data: $I_{mag}(t,q)/I_{mag}(0,q)$, where $I_{mag}(0,q)$ is determined from an average of the data before time zero. The data are plotted for a time delay $t = 205$ fs in Fig. 2b. For our FePt model we calculate the same quantity: $I_{mag}(t,q)/I_{mag}(0,q)$. To match the model data to the experimental data the core magnetization $M_{core}$, the shell magnetization $M_{shell}$ and the width of the shell region $d_{shell}$ are adjusted. The fitting is preformed independently for multiple time delays through the FePt demagnetization process. For e.g., Fig. 2b shows a good agreement between the experimental data and the fit. The model parameters extracted from fitting the experimental data are shown in Fig. 3a &

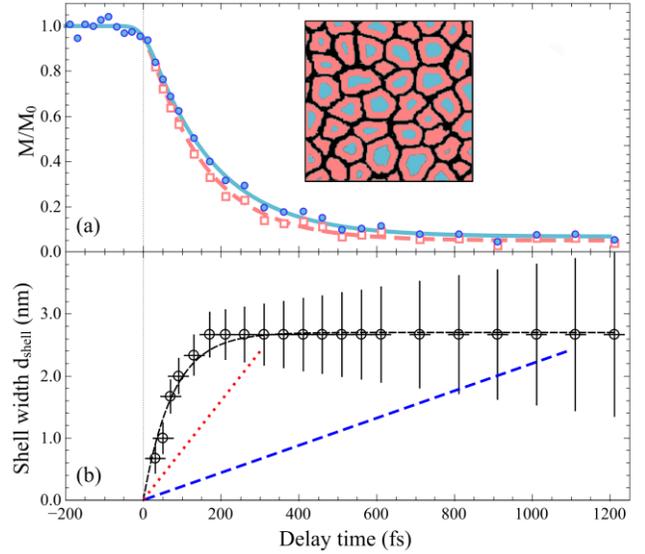

FIG. 3. Fitting parameters from the shell demagnetization model of X-ray scattering that provide best fits to these scattering data vs. time. (a) $M_{shell}$ (open squares) and $M_{core}$ (filled circles), showing demagnetization times of 135±5 and 162±4 fs respectively. Inset: image of the FePt grain modeled with partially shell demagnetization (black refers to carbon, blue to the grain core, and red the grain shell). (b) The shell thickness vs. time. The back line is an exponential fit. Magnon and phonon velocities are shown as red dotted line and blue dashed lines respectively.

3b. The extracted parameters that characterize the nanoparticle magnetization are now examined.

Both the core and shell magnetizations are observed to follow exponential (Fig. 3a, blue and red resp.). The core region magnetization shows an exponential decay constant of $\tau_{core} = 162 \pm 4$ fs while the shell magnetization shows a faster delay rate of $\tau_{shell} = 135 \pm 5$ fs. The results show that following laser excitation the shell region of the nanoparticle demagnetizes faster than the core. The difference in core and shell magnetizations ($\Delta m_{shell}$) is maximal at approximately 200 fs, from which time it begins to reduce (Fig. 4a). At time > 1ps a small, nearly constant difference between core and shell regions persists.

The thickness of the shell region can be further investigated using the comparison of data and model. Physically the thickness of the shell is encoded into the $q$ value where $I_{mag}(q)$ begins to reduce from the homogenously magnetized case. The values extracted for the shell width $d_{shell}$ are plotted in Fig. 3b. The shell width is observed to grow exponentially for the first 200 fs up to a width of 2.6 nm, from where it is observed to stay nearly constant. The stable width is close to half the width predicted for domain walls in FePt [31]. Non-magnetic surface layers have been observed in FePt nanoparticles [32] and Pt segregation has been shown to occur within a few atomic layers of the FePt surface [33]. However, while effects may contribute to the shell formation, they occur only in the first 0.5 nm from the surface. We note that any substructure within the shell region is not accessible due to the limited $q$ range of the experiment. In the following sections we will discuss the formation of this demagnetization shell and the mechanisms driving its formation.

The formation and propagation of the shell-demagnetized region is highly reminiscent of a transport phenomenon. Laser heating of FePt has been shown to form strain waves that propagate from the surface into the material [34]. Such lattice shockwaves travel at the speed of sound, which is 2200 ms$^{-1}$ along the $a,b$ lattice planes in FePt [35], as illustrated in Fig. 3b by the dashed blue line; this velocity is nearly an order of magnitude *slower* than the shell propagation. We therefore can exclude lattice dynamics as the mechanism of the shell formation. Similarly, the maximum velocity of magnetic spin waves in FePt has been determined from neutron scattering data to be 8 nm/ps (illustrated on Fig. 3b by the dotted red line) [36]. This velocity is less than a third of the initial velocity observed in the formation of the shell region. The only mechanism capable of driving magnetic transport above the spin-wave velocity is electron transport of a magnetic spin current [2,4,37,38]. We conclude that the shell region propagates with the velocity of ballistic and superdiffusive spin-flipped electrons away from the nanoparticles boundary.

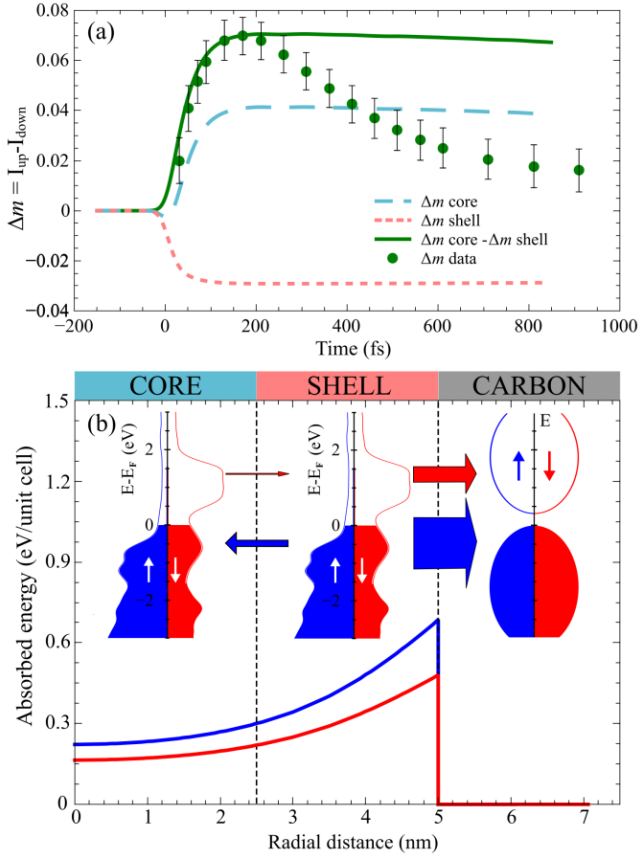

FIG. 4: Results of theoretical modelling of magnetic transport in an FePt nanoparticle following laser excitation. (a) Computed spin transport contribution to the magnetization change $\Delta m$ in the core and shell regions as a function of time, compared to the measured data (green dots). (b) The energy absorbed per FePt unit cell for spin-majority (blue) and spin-minority (red curve) electrons is plotted as a function of radial distance from the center of a 10 nm diameter FePt particle. The inset shows schematically the partial density of states of FePt and of carbon. Arrows show the spin current magnitudes (width) and directions across the core–shell and shell–carbon interfaces for majority (blue) and minority spins (red).

To further understand the dynamics and development of the shell region we quantify the magnetic moment lost in the shell region per nanoparticle $\Delta m_{shell}$. This is done by fitting the extracted shell parameters in Fig. 3a & 3b and calculating the change for the average nanoparticle. Here we define the change as being the difference between the core and shell magnetic moments:

$$\Delta m_{shell} = V_{np}M_{core} - (M_{core}V_{core} + M_{shell}V_{shell})$$
$$= V_{shell}(M_{core} - M_{shell}),$$

where $V_{np} = V_{core} + V_{shell}$. As pointed out earlier, these data show that $\Delta m_{shell}$ peaks at approximately 200 fs and then starts to decrease (Fig. 4a). The change in magnetization is associated with a spin current leaving the region, which can be quantified by taking the time derivative $\frac{d}{dt}\Delta m_{shell}$. This measurement of the change in magnetization in the volume shows that the spin current has a rapid onset (below the time resolution of the experiment) and reaches a maximum in under 40 fs.

To understand the formation of the demagnetized shell we conduct spin-transport calculations, using the superdiffusive spin current model [2] for a 10 nm FePt nanoparticle surrounded by carbon. The original model was developed to describe layered structures [2,23], which however are not present here. Therefore, we have employed the Particle In Cell (PIC) method [39] to compute the superdiffusive motion of excited electrons in general nanoscale geometries such as nanoparticles.

To start with, we use *ab initio* calculated spin-dependent refractive indices of FePt to compute the spin-dependent optical absorption, using the electromagnetic field solver Comsol 5.3; the computed spin-dependent absorptions, shown in Fig. 4b (red and blue curves), serve as an input for the excited electron profiles in the spin-transport model. In addition, we performed *ab initio* calculations of the energy and spin dependent hot electron velocities in FePt and we found that these are roughly 5 times less than those in Fe for both spin channels. The carbon interface is treated as an energy barrier, with $E_B = 0.250$ eV [20]. We define further the core region as $0 \leq r < 2.5$ nm and the shell region as $2.5 < r < 5$ nm.

Our calculations of the superdiffusive contribution to the core and shell demagnetizations are shown in Fig. 4a. The computed net demagnetization difference $\Delta m_{shell}(t)$ caused by superdiffusive transport for an average excitation of 0.63 electrons/atom (solid green line) is compared to the measured $\Delta m_{shell}$ (green circles). The superdiffusive transport contribution qualitatively reproduces the experimental observations, in terms of onset time and contribution size. We note however, first, that the experimental observed decay ($t > 200$ fs) is not observed in the calculations. This is attributed to the thermal diffusion of the holes and local spin flips, such as electron-magnon processes, that are not taken into account in our model. Second, to have the same maximum $\Delta m$ value as in the experiment, we assume a 1.4 times larger excitation density than that obtained from *ab initio* bulk FePt calculations.

To understand the role of spin transport in the formation of the shell demagnetization region we examine the net spin transport across the core-shell and shell-carbon interfaces in the first 400 fs of the simulation (see Fig. 4b). The transport is computed to be dominated by a loss of spins from the shell region through the carbon barrier. A spin current of 0.094 majority spins/unit cell and 0.030 minority spins/unit cell leave the shell region into the carbon matrix, giving a net spin current of 0.064 spins/unit cell. A smaller spin current, of 0.013 majority spins/unit cell and -0.003 minority spins/unit cell, is observed to flow from the shell region into the nanoparticle core. This gives a net spin current of 0.016 spins/unit cell leaving the shell region and acting to increase the difference in magnetizations between core and shell regions. Therefore, we find that the shell formation is primarily driven by spin transport into the surrounding carbon matrix.

It has to be stressed that the transport contribution to the nanoparticle demagnetization is not dominant in FePt. While electronic transport of spin affects predominantly the magnetization at the boundary of the nanoparticles, this accounts for only ~6% of the loss of magnetization. The X-ray scattering data shows unambiguously that spin and orbital angular momentum are lost throughout the nanoparticle via a local transfer to an angular momentum reservoir. We attribute this tentatively to ultrafast spin–lattice transfer.

In conclusion, X-ray scattering data shows that during laser induced demagnetization of FePt nanoparticles a region of increased demagnetization develops at the edges of FePt nanoparticles, which proceeds inwards with high velocity. Superdiffusive spin-transport simulations show that the formation and propagation velocity of the shell region is caused by ultrafast spin transport from the nanoparticle into the surrounding carbon. Our results highlight a new possible mechanism of how magnetic boundaries can form and propagate in ferromagnetic nanosystems at *supermagnonic* velocities. Knowledge of such mechanism, and the ability to utilize it, will be of importance for achieving controllable spin dynamics in the nm-fs domain.

This work was supported by U.S. Department of Energy, Office of Science under Award Number 0000231415. Work at SIMES and the use of the Linac Coherent Light Source (LCLS), SLAC National Accelerator Laboratory, was supported by the U.S. Department of Energy, Office of Science, Office of Basic Energy Sciences under Contract No. DE-AC02-76SF00515. Work at UU was supported by the Swedish Research Council (VR), the K. and A. Wallenberg Foundation (grant No. 2015.0060), the European Union's Horizon2020 Research and Innovation Programme under Grant Agreement No. 737709, and the Swedish National Infrastructure for Computing (SNIC). Work at NIMS was supported by JSPS KAKENHI Grant No. 18H03787. L.L.G. would like to thank the Volkswagen-Stiftung for the financial support through the Peter-Paul-Ewald Fellowship. We thank P. Maldonado for helpful comments.

*alexhmr@slac.stanford.edu
#hermann.durr@physics.uu.se